\documentclass[preprint,12pt]{elsarticle}
\usepackage{amsmath,graphicx}
\usepackage{bm}

\journal{Journal of Solid State Chemistry}

\begin{document}

\begin{frontmatter}

\title{A Beginner's Guide to the Modern Theory of Polarization}
\author{Nicola~A.~Spaldin}
\address{Materials Theory, ETH Zurich, Wolfgang-Pauli-Strasse 27, 8093 Z\"{u}rich, Switzerland}

\begin{abstract}
The so-called {\it Modern Theory of Polarization}, which rigorously defines the
spontaneous polarization of a periodic solid and provides a route for its computation
in electronic structure codes through the Berry phase, is introduced in a simple
qualitative discussion. 
\end{abstract}


\end{frontmatter}

\section{Introduction}

The concept of electric dipole moment is central in the theory of electrostatics, 
particularly in describing the response of systems to applied electric fields. For
finite systems such as molecules it poses no conceptual or practical problems. 
In the ionic limit the dipole moment, $d$, of a collection of charges, $q_i$, 
at positions $\mathbf{r}_i$ is defined as
\begin{equation}
d = \Sigma_i q_i \mathbf{r}_i  \quad ;
\end{equation}
for the case of a continuous charge density, $e n(\mathbf{r})$ (where $e$ is the electronic
charge and $n(\mathbf{r})$ is the number density) this expression is straightforwardly
extended to
\begin{equation}
d = \int e n(\mathbf{r}) \mathbf{r} d\mathbf{r} \quad .
\end{equation}
Provided that the molecule or cluster carries no net charge these expressions are well
defined, can be straightforwardly evaluated and yield results -- for example for the direction
of the dipole moment -- that are consistent with our intuitive understanding.

Things apparently start to turn to custard, however, when we try to extend this simple
reasoning to bulk solids. The usual way to define intrinsic quantities in macroscopic
systems is to introduce the property per unit volume or mass. For example the magnetization
is the magnetic moment per unit volume, and the bulk
analogue to the electric dipole moment, the electric polarization, should be represented by
the electric dipole moment per unit volume. The relevant quantity is then evaluated within
a small repeat unit -- the unit cell -- of the solid, and normalized with the
volume of the chosen unit cell. The problem with this simple method in the case of electric 
polarization can be understood in the simple one-dimensional cartoon of Figure~\ref{1Dchain}: 
Without performing any calculations, we can see that the two equally valid unit cells 
shown with dashed lines have completely {\it opposite} orientations of the polarization!

This difficulty led to tremendous confusion in the field, with discussions as fundamental
as whether the polarization (and related quantities such as the piezoelectric response) 
could be considered as intrinsic properties in bulk solids, or are in fact determined
by details of the surface termination. Thankfully the confusion was resolved around 20 years ago with
the introduction of the so-called {\it Modern theory of polarization}. This very elegant
theory showed that {\it changes} in polarization are in fact rigorously defined, can
be calculated quantum mechanically using electronic structure methods, and correspond
to experimentally measurable observables.

The purpose of this article is to introduce in the simplest possible terms the apparent
difficulties associated with defining polarization in bulk solids, and the solutions
provided by the modern theory. It is motivated by my having explained these concepts
repeatedly to many and diverse students ranging from experimentalists with a casual interest in
understanding obscure theory papers to beginning hard-core theoretical solid-state physicists 
and quantum chemists. This article in
no way intends to substitute for the elegant early papers on the topic, nor the
subsequent detailed and rigorous review papers which are referenced throughout. Indeed
I hope that this informal introduction provides sufficient background for the reader
to tackle these excellent articles without intimidation. 

\section{Bulk periodicity, the polarization lattice and the polarization quantum}

\begin{figure}[ht]
 \centering
 \includegraphics[width=1.0\columnwidth]{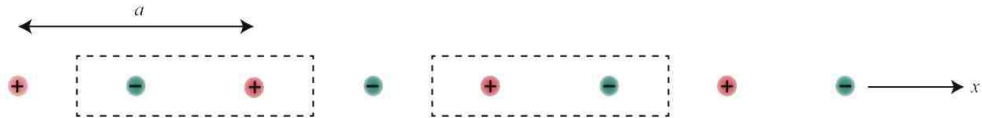}
 \caption{One-dimensional chain of alternating anions and cations, spaced a distance
$a/2$ apart, where $a$ is the lattice constant. The dashed lines indicate two 
representative unit cells which are used in the text for calculation of the polarization.}
 \label{1Dchain}
\end{figure}

We begin by reconciling the different values for polarization obtained for the different 
choices of unit cells in Fig.~\ref{1Dchain} by introducing a formal concept that at first
sight is even more confusing -- that is the {\it multi-valuedness} of the bulk polarization. 
We will show, however, that a multi-valued polarization is a natural consequence of the periodicity
in a bulk solid, and hopefully that it is actually not so frightening. We will see, in fact 
that {\it changes} in polarization -- which are the quantities that are anyway measured in 
experiments -- are single valued and well defined, and we can once again sleep without anxiety.  

We take the simplest possible example of a one-dimensional chain of singly charged
alternating anions and cations -- the closest real-life analogue would be rock-salt
structure sodium chloride in just one direction. Look at Figure~\ref{1Dchain} which
shows such a chain with the atoms spaced a distance $a/2$ apart so that the lattice
constant is $a$. The first thing to notice is that all of the ions are centers of
inversion symmetry: If I sit on any ion and look to the left, then to the right I see 
no difference. So by definition this lattice is non-polar. 

Now let's work out the polarization by calculating the dipole moment {\it per unit length}
(the definition in three dimensions is dipole moment per unit volume) using in turn 
the two unit cells 
shown as the dashed rectangles to compute the local dipole moment. First, the cell on the left.
Taking the left edge of the shell as the origin ($x=0$), we have an ion with charge -1 at
position $a/4$, and an ion with charge +1 at position $3a/4$. So the polarization, or dipole moment
per unit length is
\begin{eqnarray}
p & = & \frac{1}{a} \sum_i q_i x_i \nonumber\\
  & = & \frac{1}{a} \left(-1 \times \frac{a}{4} +1 \times \frac{3a}{4} \right) \nonumber\\
  & = & \frac{1}{a} \frac{2a}{4} \nonumber\\
  & = & \frac{1}{2} 
\end{eqnarray}
in units of $|e|$ per unit length.
Immediately we have an apparent problem: Using this method, our non-polar chain has a non-zero 
polarization.

I am afraid that things will get worse before they get better. Next, let's do the same
exercise using the right-most unit cell. Again taking the left edge of the unit cell as
the origin, this time there is a positively charged ion at position $a/4$, and a negatively
charged ion at $3a/4$. So
\begin{eqnarray}
p & = & \frac{1}{a} \left(+1 \times \frac{a}{4} -1 \times \frac{3a}{4} \right) \nonumber\\
  & = & \frac{1}{a} \times -\frac{2a}{4} \nonumber\\
  & = & -\frac{1}{2} \quad .
\end{eqnarray}
Again a non-zero value, and this time {\it different} from the value we obtained using the
other, equally valid unit cell, by an amount $a$. 

So what is going on here, and how can we connect it to physical reality? Well, if we were to 
repeat this exercise with many choices of unit cell (convince yourself by choosing a couple
of arbitrary unit cells and giving it a try!), we would obtain many values of polarization, 
with each value differing from the original value by an integer. 
We call this collection of polarization values the
{\it polarization lattice}. In this case it is $..., -5/2, -3/2, -1/2, 1/2, 3/2, 5/2 ...$.
Notice that the lattice of polarization values is symmetric about the origin. In fact this is the
signature of a non-polar structure: The polarization lattice may or may not contain zero as one
of its elements, but it must be centrosymmetric around zero. 

Now what is the significance of the spacing (in this case 1) between the allowed values? 
Well, imagine removing an electron from one of the anions in the lattice (leaving a neutral atom)
and moving it by one unit cell to put it on the next anion to the right. Because of the periodic 
boundary conditions of the infinite lattice, the next anion simultaneously has it's electron 
removed and moved one unit cell to the right, and so it is able to accept the incoming electron 
and appear unchanged at the end of the process. There has been no change in the physics of the
system resulting from the relocation of the electrons by one unit cell to the right. But what
has happened to the polarization? Well, in each unit cell a charge of $-1$ has moved a distance $a$,
changing the dipole moment by $-a$ and the polarization by -1. We can clearly perform this thought 
experiment any number of times,
and in either direction, changing the polarization by any integer without changing the
physical system! We call the value of polarization resulting from moving one electron by one unit cell
the {\it polarization quantum}, $P_q$. In one dimension it is equal to the lattice constant divided
by the length of the unit cell, which is just an integer (in units of the
electronic charge per unit length). Going back to the polarization lattice of our non-polar
chain, we see that it's polarization values correspond to half-polarization quanta. In fact
all non-polar systems have polarization lattices of either $0 \pm nP_q$ or $\frac{P_q}{2} \pm nP_q$.

\begin{figure}[ht]
 \centering
 \includegraphics[width=1.0\columnwidth]{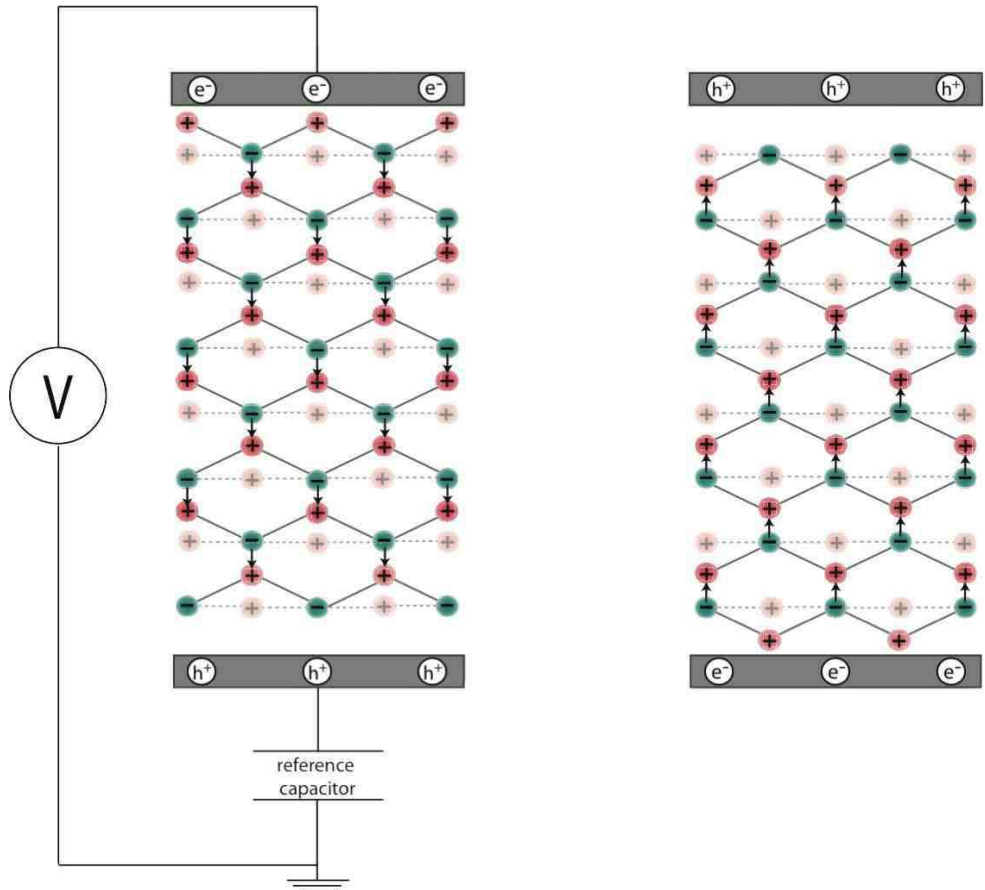}
 \caption{Schematic of the Sawyer-Tower method of measuring ferroelectric polarization. The
material on the left is polarized in the  up direction and its surface charge is screened by
electrons in the upper electrode (grey) and holes in the lower electrode. When the polarization
is switched (right), electrons and holes flow through the external circuit to screen the new
opposite surface charges, and are counted by comparing the voltage across the material with 
that across a reference capacitor.}
 \label{Sawyer-Tower}
\end{figure}

If this all seems too esoteric, please bear with me for one more paragraph by which time I 
hope things should
start making sense. First, let's think about how we measure electrical polarization, and what 
a reported measured polarization really means.
Look at Figure~\ref{Sawyer-Tower} -- this is a cartoon of a standard way of measuring the electrical 
polarization using a so-called Sawyer-Tower circuit. On the left the material has become polarized 
in the up direction as a result of the cation sub-lattice displacing upwards relative to the 
anion sub-lattice. This could happen, for example during a ferroelectric phase transition with
an external electric field applied in the up direction (the light colored cations with the 
dashed-line bonds indicate their positions in the high-symmetry, paraelectric structure). Electrons 
accumulate at the upper electrode, and holes (or a depletion of electrons) at the lower electrode 
in order to screen the surface charge resulting from the ionic displacements. In fact, on each
electrode, the accumulated charge per unit area is exactly equal to the polarization of the
sample. So if we could measure the amount of charge accumulation we would have a direct measure 
of the polarization. But how can we do this? Well, next, imagine
reversing the orientation of the polarization -- for example by applying an 
external electric field in the down direction -- to reach the configuration on the right.
Now electrons accumulate at the lower electrode and holes at the upper electrode to achieve
the screening. They achieve this by flowing through the external circuit connecting the two
electrodes, where they can be counted by comparing the voltage across the series reference capacitor
then using $Q=CV$! The amount of charge per unit area of electrode that flows during the
transition is equal to the change in polarization between the up- and down-polarized states; 
the ``absolute'' value of polarization which is reported is half of this number.

\begin{figure}[ht]
 \centering
 \includegraphics[width=1.0\columnwidth]{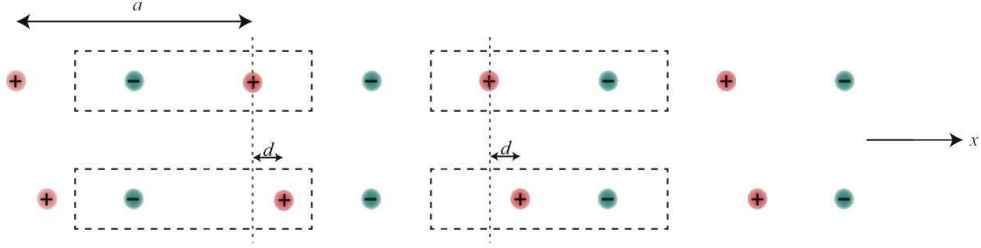}
 \caption{The upper panel reproduces the one-dimensional chain of alternating anions and 
cations of Fig.~\ref{1Dchain}. In the lower panel,
the cations are displaced to the right by a distance $d$ relative to the anions, with 
the vertical dotted lines indicating their original positions.}
 \label{1Dchain_polar}
\end{figure}

Now, bearing in mind that what is measured in an experiment is a {\it change} in polarization,
let's go back to our cartoon one-dimensional model and make some sense out of this 
multi-valuedness business. In the upper part of Figure~\ref{1Dchain_polar} we reproduce the 
non-polar one-dimensional chain of Fig.~\ref{1Dchain}, and below it we show a similar chain
in which the cations have been displaced by a distance $d$ relative to the anions in the 
manner of a ferroelectric distortion to create a polar system. Let's repeat our earlier
exercise of calculating the polarization using the two unit cells shown as the dashed
rectangles. 

\clearpage

In the left hand case,
\begin{eqnarray}
p = \frac{d}{a} & = & \frac{1}{a} \sum_i q_i x_i \nonumber\\
                & = & \frac{1}{a} \left(-1 \times \frac{a}{4} +1 \times (\frac{3a}{4} +d) \right) \nonumber\\
                & = & \frac{1}{2} + \frac{d}{a} 
\end{eqnarray}
and in the right hand case,
\begin{eqnarray}
p & = & \frac{1}{a} \left(+1 \times (\frac{a}{4} + d) -1 \times \frac{3a}{4} \right) \nonumber\\
  & = & -\frac{1}{2} + \frac{d}{a} \quad .
\end{eqnarray}
Again the two answers are different, but this time that doesn't worry us, because we recognize
that they differ by exactly one polarization quantum. Next comes the key point: Let's 
calculate the {\it change} in polarization between the polar and non-polar chains using each
unit cell as our basis. First for the cell on the left,
\begin{equation}
\delta p = (\frac{1}{2} + \frac{d}{a} ) -\frac{1}{2} = \frac{d}{a}
\end{equation}
and for the cell on the right
\begin{equation}
\delta p = -(-\frac{1}{2} + \frac{d}{a} ) - (-\frac{1}{2}) = \frac{d}{a} \quad.
\end{equation}
In both cases the change in polarization between polar and non-polar chains is the same. In
fact this would have been the case whatever unit cell we had chosen to make the calculation.
So, while the absolute value of polarization in a bulk, periodic system, is multivalued,
the change in polarization -- which remember is the quantity that can be measured in an
experiment -- is single valued and well defined. Phew.

\begin{figure}[ht]
 \centering
 \includegraphics[width=0.7\columnwidth]{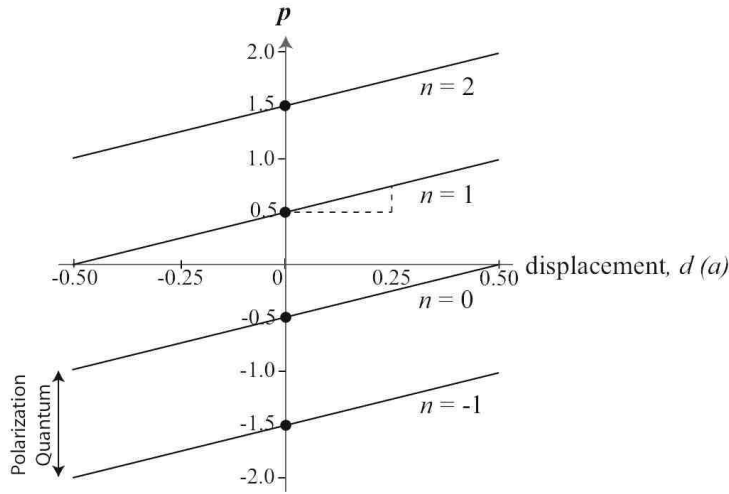}
 \caption{Polarization as a function of the displacement, $d$, of the cations in the 
1D chain of Figure~\ref{1Dchain_polar}. The polarization lattice is zero-centered, and
the branches are separated by the polarization quantum. Notice that the branches of the
lattice run exactly parallel to each other, so that differences in polarization along 
each branch for the same displacement are identical.}
 \label{p_of_d}
\end{figure}

Just to really drive the point home, in Fig.~\ref{p_of_d} we plot the polarization of
the ideal one-dimensional ionic chain as a function of the displacement of 
the cations (as a fraction of the lattice constant) from their non-polar positions. As we calculated 
earlier, for zero displacement the polarization lattice is centrosymmetric and consists 
of all half-integer values (black circles). As the displacement increases, the polarization increases
linearly and by the same amount along each branch of the polarization lattice (labeled
by $n = -1, 0, 1$ etc.) The branches are always separated from each other by the same amount, the
polarization quantum, which is equal to
$1$ in this case. The dashed lines on the $n=1$ branch show that for a displacement
of 0.25$a$, the polarization increases from 0.5 to 0.75, and so the change in polarization
is 0.25. If the displacement is increased artificially to 0.5 -- that is half of the 
unit cell -- the ions end up on top of each other; in our thought experiment this causes
the polarization to jump between branches of the polarization lattice, although in practice 
we would have achived nuclear fusion which would likely dominate the physics. 

\subsection{Extension to three dimensions}

The one-dimensional example that we chose here for simplicity is not entirely without
physical relevance; for example ferroelectric or polar polymers closely resemble one-dimensional
chains. Such an application is discussed in Ref.~\cite{Kudin/Car/Resta_2:2007}, along with
an excellent analysis of the development of an infinite chain from a finite one. In most cases,
however, we are interested in three-dimensional systems, and fortunately the extension to three
dimensions is straightforward: A polarization lattice can now be defined along all three 
lattice vector directions, with the polarization quantum equal to 
\begin{equation}
P_i = \frac{1}{\Omega} e R_i
\end{equation}
Here $e$ is the electronic charge, $R_i$ is the $i$th lattice vector, and $\Omega$ is the unit 
cell volume. Note that in non-magnetic systems, the polarization quantum is usually multiplied by an 
additional factor of two because the up- and down-spin electrons are equivalent, and shifting
an up-spin electron by a  lattice vector also shifts the corresponding down-spin electron. 
Polarization values along cartesian coordinates, for example, can then be readily obtained
using the appropriate linear transformation.

\section{Wannier representation and Berry Phase}
\label{section_WR_and_BP}

In the previous section we discussed the multivaluedness of the polarization in a bulk
periodic solid and reconciled it with what can be measured experimentally for the simple 
example of an array of ions. Of course in a real solid,
there is (thankfully) more chemistry to take care of. In this section we will first 
explain how this chemistry can be incorporated rather simply by extending the ionic 
model through the method of Wannier functions. (A similar approach is followed in Ref.
~\cite{Kudin/Car/Resta_1:2007}, where an algorithm is developed that is particularly 
suited to localized-basis quantum chemistry codes.) Once we are comfortable with this 
conceptually we will move on to the real meat of the modern theory of polarization -- 
the Berry phase method.

Remember that the Wannier function, $w_n(\mathbf{r})$, in unit cell $\mathbf{R}$ associated 
with band $n$ is defined as
\begin{eqnarray}
w_n(\mathbf{r}-\mathbf{R}) & = & 
\frac{\Omega}{(2\pi)^3} \int_{BZ} d^3 \mathbf{k} e^{-i \mathbf{k.R}} \Psi_{n \mathbf{k}}(\mathbf{r})\nonumber \\
& = & 
\frac{\Omega}{(2\pi)^3} \int_{BZ} d^3 \mathbf{k} e^{i \mathbf{k.(r} - \mathbf{R})} u_{n \mathbf{k}}(\mathbf{r})
\end{eqnarray}
where $\Psi_{n \mathbf{k}}(\mathbf{r})= e^{i\mathbf{k.r}} u_{n\mathbf{k}}(\mathbf{r})$
are the Bloch functions, written as usual in terms of the cell-periodic part, $u_{n\mathbf{k}}(\mathbf{r})$.
Here $\Omega$ is the unit cell volume, and the integral is over the Brillouin zone.

Unlike the Bloch functions which are delocalized in space, the Wannier functions are localized.
As a result they are often used in visualization of chemical bonding, as well as for basis sets
in electronic structure calculations, where their minimal overlap can lead to favorable scaling
with system size. They are relevant here, because their localized nature provides a convenient
atomic-like description of the charge density in a solid: While we know in reality that the
charge density in a solid is a continuous function, the localized picture will allow us to continue
to calculate dipole moments by summing over charges multiplied by positions. 

\begin{figure}[ht]
 \centering
 \includegraphics[width=1.0\columnwidth]{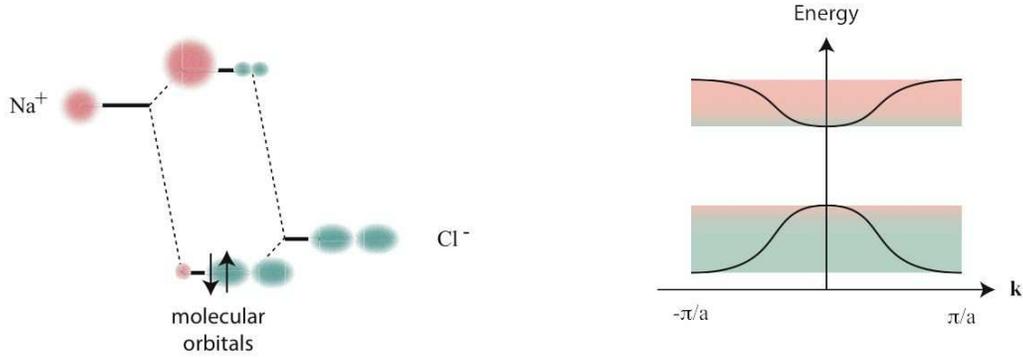}
 \caption{Left: The molecular orbitals formed in an Na-Cl ``molecule''. Right: Band
structure of a one-dimensional Na-Cl chain. The valence band is derived from Cl-like 
molecular orbitals each containing two electrons, and is fully occupied; the Na-like 
conduction band is empty. In both cases pink represents Na-derived states and green 
Cl-derived states.
}
 \label{NaCl_BandStructure}
\end{figure}

\begin{figure}[ht]
 \centering
 \includegraphics[width=1.0\columnwidth]{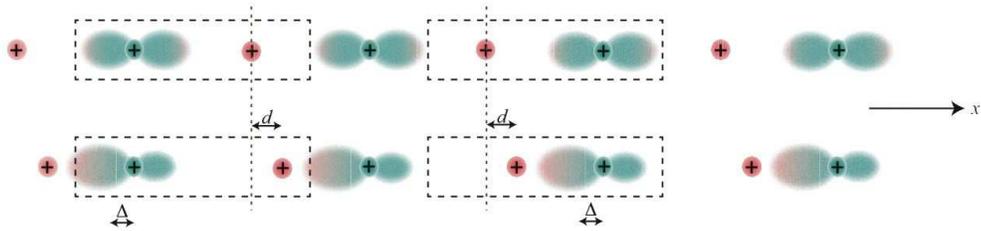}
 \caption{One-dimensional chain of alternating cations (pink postively charged ion cores) 
and anions (green positively charged ion cores with their associated negatively charged 
valence electron cloud). The dimensions and dashed unit cells are as in Figure~\ref{1Dchain}.}
 \label{1Dchain_WF}
\end{figure}

Let's go back to our 1D chain, and relax the constraint that it is composed of point charge
ions to give it some chemistry. If it's helpful you could think of it as say a chain of
Na$^{+}$ cations alternating with Cl$^{-}$ anions. In the following figures we associate pink with
Na ions or electrons, and green with Cl ions or electrons. In Figure~\ref{NaCl_BandStructure} (left) we 
show the molecular orbitals that would form between two such ions in an Na-Cl ``molecule'' -- 
the lower energy, bonding orbital is occupied by two electrons and more localized on the $p$ 
orbital of the anion, and the higher energy, 
antibonding orbital is empty and consists primarily of cation $s$ character. The corresponding band 
structure cartoon is shown to the right; you can derive the dispersion using simple 
linear-combination-of-atomic-orbitals (LCAO) methods; see for example the book by Cox \cite{Cox:Book}.
In Figure~\ref{1Dchain_WF} we show a cartoon
of our 1D chain again, but this time we have separated out the charge on the ions (all of
which are +1, and which we continue to treat as point charges) from the charge on the electrons
which are spread through the system, but piled up more on the anions than the cations.
The blobs around the anions illustrate what we might expect the Wannier functions of the
occupied band to look like, with each Wannier function containing two electrons. The character
of the Wannier function is mostly Cl $p$-like, with a little bit of Na $s$ character, indicated
by the slight pink tinge on the edges.
Note that if we consider both of the electrons in each Wannier function to be associated with the
Cl ion, then the formal charge on the Cl is +1 (the ionic charge) -2 = -1, and that on the Na
ion is +1 +0 = +1, and we recover our simple ionic model of Figs.~\ref{1Dchain} and \ref{1Dchain_polar}. 

How should we now calculate the polarization of the chain? We would like again
to reduce our polarization integral again to a sum over localized charges multiplied by their 
positions. This is straightforward for the ions which we are still treating as point
charges. For the electrons, it turns out that this procedure will work too. Since the 
Wannier functions are localized, we work out the average position of the electrons
in the Wannier function, and treat them all as sitting at that point. This ``position''
of the Wannier function is called the {\it Wannier center}, $\overline{\mathbf{r}}_n$. 
The Wannier center associated with band $n$ is defined to be the 
expectation value of the position operator $\mathbf{r}$ for Wannier function $w_n(\mathbf{r})$:
\begin{equation}
\overline{\mathbf{r}}_n=\int w_n^*(\mathbf{r}) \mathbf{r} w_n(\mathbf{r}) d^3 \mathbf{r}
\label{Eq_Wannier_Center_r}
\end{equation}
Later we will find it useful to rewrite this expression in terms of the periodic
cell functions using the momentum representation of the position operator
$\mathbf{r} = -i \frac{\partial}{\partial \mathbf{k}}$:
\begin{equation}
\overline{\mathbf{r}}_n=i\frac{\Omega}{(2\pi)^3} \int_{BZ}d^3\mathbf{k}
e^{-i \mathbf{k}.\mathbf{R}} \left< u_{n \mathbf{k}} | \frac{\partial u_{n \mathbf{k}}}
{\partial \mathbf{k}} \right> 
\label{Eq_Wannier_Center_k}
\end{equation}
You can spend your next free Sunday morning showing that Eqns. ~\ref{Eq_Wannier_Center_r} and 
~\ref{Eq_Wannier_Center_k} are equivalent, take
my word for it, or follow the derivation by Blount in Ref. ~\cite{Blount:1962}

With this concept of the Wannier center, the expression for polarization that we used 
previously for the ionic chain extends simply to a sum over the contribution from
the point charge ions, plus a sum over the electronic charges centered at the Wannier 
centers of each occupied Wannier function, $n$:
\begin{equation}
p  =  \frac{1}{a} \left(\sum_i (q_i x_i)^{ions} + \sum_n^{occ} (q_n \overline{\mathbf{r}}_{n})^{WFs} \right) \\
\label{PofWF}
\end{equation}
Let's try it for the case of the left-hand unit cell in our 1D chain. In the non-polar case,
we can see by symmetry that the Wannier center is at the same position as the green anion; 
remember now also that the 
charge on all of the ions is +1, and that each Wannier function contains two electrons. So
the dipole moment per unit length in the left unit cell is
\begin{eqnarray}
p  & = & \frac{1}{a} \left(+1 \times \frac{a}{4} +1 \times \frac{3a}{4} + -2 \times \frac{a}{4} \right) \nonumber\\
  & = & \frac{1}{a} \frac{2a}{4}\nonumber\\
  & = & \frac{1}{2} \quad .
\end{eqnarray}
The same result as we obtained previously! This is as expected -- the allowed values of
the polarization lattice for a centrosymmetric structure are dictated by the symmetry of
the crystal and the ionic charges, and are not modified by factors such as the details of the chemical
bonding within the material.

Now let's think about the off-centered case, in the lower part of Fig.~\ref{1Dchain_WF}.
As before, the cations have moved a distance $d$ to the right, but this time the Wannier
centers have also moved -- by a distance $\Delta$ say -- to the left. This occurs as the 
chemical bond between the near neighbor anion-cation pairs becomes stronger, and develops more
cation $s$ character, whereas that between the distant neighbor pairs weakens; you can think of
it as a flow of electrons from the anion (which previously had all of the valence electrons)
towards the cation in the process of covalent bond formation. Let's see what this additional
covalency does to the polarization:
\begin{eqnarray}
p  & = & \frac{1}{a} \left(+1 \frac{a}{4} +1 (\frac{3a}{4} + d) + 
                           -2 (\frac{a}{4} - \Delta) \right) \nonumber\\
  & = & \frac{1}{a} (\frac{2a}{4} + d + 2\Delta)\nonumber\\
  & = & \frac{1}{2} +\frac{d}{a} +2\frac{\Delta}{a}\quad .
\end{eqnarray}
Compared to the purely ionic case, the polarization has increased by an amount
$\frac{2\Delta}{a}$. This is because, in addition to the positively charged cation
moving to the right (along the positive $x$ axis) some negatively charged electron
density has moved to the left (along the negative $x$ axis). This results in a
larger {\it effective} displacement of positive charge along $+x$ and a larger 
polarization. We'll return to this picture later when we discuss the concept
of the Born effective charge. 

Let's summarize the discussion so far before we go on to formalize it mathematically.
Using our Wannier function picture we can continue to write our polarization as the 
sum over the charges times their positions. We include both the contribution from the
postively charged ion cores, and the contribution from the negatively charged valence electrons,
and we take the ``position'' of each valence electron to be its Wannier center.
Had we repeated our analysis for the right-hand unit cell, we would have seen that,
as in the purely ionic model, the polarization is multi-valued, but the difference
in polarization for example between a centrosymmetric and polar structure, is well-defined, 
and corresponds to the experimentally measurable spontaneous polarization.

Now we will derive the formal mathematical expression for the spontaneous polarization 
$\delta p$ in the Wannier representation. Since we already have an expression for the Wannier
centers this is going to be rather painless. Remember that the spontaneous polarization
is the difference in polarization, on the same branch of the polarization lattice, between 
the final, polarized and initial, unpolarized states. Using 
$p = \frac{1}{\Omega} \sum_i q_i \mathbf{r}_i$ for
the ionic part, and Eqn.~\ref{Eq_Wannier_Center_k} for the Wannier centers, we obtain:
\begin{eqnarray}
\delta p & = & p^f - p^0 \nonumber \\
         & = & \frac{1}{\Omega} \sum_i \left[ q_i^f \mathbf{r}_i^f - q_i^0 \mathbf{r}_i^0 \right] \nonumber \\
         & - & \frac{2ie}{(2\pi)^3} \sum_n^{occ} \left[ \int_{BZ} d^3 \mathbf{k} e^{-i\mathbf{k.R}} 
 \left< u_{n \mathbf{k}}^f| \frac{\partial u_{n \mathbf{k}}^f}{\partial \mathbf{k}} \right>
-\left< u_{n \mathbf{k}}^0| \frac{\partial u_{n \mathbf{k}}^0}{\partial \mathbf{k}} \right>
\right] 
\label{P_BP}
\end{eqnarray}
where $f$ and $0$ indicate the final (polar) and intial (high symmetry) positions/wavefunctions.
Since the wavefunctions, at least at the Kohn-Sham level, are a direct output of standard
electronic structure codes, Eqn.~\ref{P_BP} can be used to evaluate the polarization with
only a small extension to a standard density functional theory code. (A rigorous extension to correlated,
many-body wavefunctions also exists, see for example Refs. \cite{Resta:1999} and \cite{Resta:1998}.)
Notice of course, that
the issues discussed earlier about multivaluedness of the polarization and the polarization
lattice persist here, and in taking the difference in Eqn.~\ref{P_BP} one must be careful
to remain on the same branch of the polarization lattice. 

If you are familiar with the concept of the {\it Berry phase} \cite{Berry:1984} and its
extension to periodic solids \cite{Zak:1989} you will recognize the integrals in Eqn.~\ref{P_BP}
to be the Berry phase developed by the wavefunction $u_{n \mathbf{k}}$ as it evolves along
the path $\mathbf{k}$. As a result, the formalism for calculating polarization using this
method is often called the {\it Berry phase theory of polarization}. Refs. 
~\cite{King-Smith/Vanderbilt:1993, King-Smith/Vanderbilt:1994, Resta:1993, Resta:1994}
are the original papers providing the detailed derivations of the Berry phase formalism, and 
excellent reviews can be found
in Refs. ~\cite{Resta:1996,Martin:Book,Resta/Vanderbilt:2007}. If you find the Berry 
phase concept too frightening, however, just stick with the Wannier function ideas, and regard
Eqn. ~\ref{P_BP} as a tool that we'll see in Section~\ref{nittygritties} allows for convenient 
computation.

\subsection{Subtlety -- gauge transformation!}
Those of you who have managed to stay awake and alert to this point might raise an objection:
Since the Bloch functions are defined only to within a phase factor, i.e. 
\begin{equation}
\Psi_{n \mathbf{k}}(\mathbf{r}) \rightarrow e^{i\phi(\mathbf{k})} \Psi_{n\mathbf{k}}(\mathbf{r})
\end{equation}
without changing any physically meaningful quantities, the Wannier functions
are not uniquely defined! As a result, the Wannier centers, which we have just seen
are crucial in defining the polarization, are also not uniquely defined. We are
saved, however, by the fact that the sum over the Wannier centers in any given
unit cell {\it is} uniquely defined, and looking again at Eqn.~\ref{PofWF} we find that
this is in fact the quantity that matters in defining the polarization.
In practice, special choices of Wannier functions are often made in calculations of the 
polarization. The so-called {\it Maximally-localized Wannier functions} in which
the phases of the Bloch functions are chosen so as to minimize the sum of the mean 
squares of the positional spread \cite{Marzari/Vanderbilt:1997} are particularly
popular. 

\section{The concept of Born Effective Charge}

At this stage I think it's appropriate to formally introduce the {\it Born effective charge},
which is a quantity that is very useful conceptually in thinking about ferroelectric 
polarization. In fact we have already seen the main idea, in Section~\ref{section_WR_and_BP},
where we saw that the polarization resulting from the displacement of an ion could be different 
from that expected by multiplying its formal charge times its displacement, in the case when
the Wannier center(s) move by a different amount than the ion cores. In fact in the example of 
Fig.~\ref{1Dchain_WF}, as the positive cations moved to the right, the Wannier centers 
shifted to the left, resulting in a larger overall polarization than we would have expected from
the formal charges alone. We say that the {\it effective charges} on the ions -- the amount
of charge that effectively contributes to the polarization during the displacement -- is larger 
than the formal charge.

This is formalized in the concept of the Born effective charge, $Z^*$, which is defined as
the change in polarization divided by the amount that an ion (or rather the periodic sub-lattice
of equivalent ions) is displaced:
\begin{equation}
Z^*_{ij} = \frac{\Omega}{e}\frac{\delta P_i}{\delta d_j} \quad.
\end{equation}
The Born effective charge is a tensor: When an ionic sublattice is displaced in direction
$i$, there is of course a change in polarization along the displacement direction, but in
addition, the polarization in perpendicular directions, $j$, can change. Turning this 
expression around we can see immediately what we have been discussing qualitatively --
that the change in polarization is determined by these {\it effective charges} times
their displacements, not by the formal charges:
\begin{equation}
\partial P_i = \frac{e}{\Omega} Z^*_{ij} \delta d_j
\end{equation}
The total polarization is then obtained by summing over the contributions from the displacements
of all sublattices.

In materials that are ferroelectric, or that are close to a ferroelectric phase transition,
the Born effective charges tend to be anomalously large, particularly on the atoms that
displace the furthest from their high symmetry to their ferroelectric positions. For 
example in the prototypical ferroelectric PbTiO$_3$, in which the formal charges are
Pb $+2$, Ti $+4$ and O $2-$, the effective charges on the ions that are active during the
ferroelectric phase transition are Pb $+3.9$, Ti $+7.1$ and O $-5.8$ \cite{Ghosez/Michenaud/Gonze:1998}.
This is consistent, with the alternative, equivalent definition of the Born effective charge as
the force induced on an ion by a uniform small electric field, $E$:
\begin{equation}
Z^*_{ij} = -e\frac{\delta F_i}{\delta E_j} \quad.
\end{equation}
In highly polarizable ferroelectrics, small electric fields generate large forces on the
ions, mediated by the anomalously large Born effective charges.

Lastly, I want to emphasize that it is important to distinguish between the Born effective
charge, which is a well-defined {\it dynamical} and {\it measurable} quantity, and the 
{\it formal, static} charge on an ion. The latter quantity, which reflects the number of 
electrons sitting at a particular ion site, depends on how you 
``count'', since there is not a unique way of deciding how to apportion the electrons 
in a chemical bond to one ion or another. While the static charge indeed indicates 
a measure of the amount of covalency in a compound, it is not a good indicator of 
ferroelectricity, which is rather indicated by a {\it change} in covalency during 
ionic displacement.

\section{A few tips on getting a Berry Phase calculation to work} 
\label{nittygritties}

Finally we describe a few of the tricks and foibles that we have learned through (sometimes) 
bitter experience 
are needed to make a Berry phase calculation of the polarization both run and give the correct answer. We try
to keep our comments general -- for the specifics of a particular code refer to the
relevant manual. 

The first step is of course to calculate the structure (if required) and self-consistent 
charge density, as in any standard total energy calculation. Of course the charge density should be 
well-converged with respect to the energy cutoff and $\mathbf{k}$-point sampling. In addition, if one 
is interested in systems such as improper ferroelectrics with small polarization values
\cite{Kimura_et_al_Nature:2003,Malashevich/Vanderbilt:2008}, 
the ionic positions must be obtained with higher-than-usual accuracy. An extra subtlety is
to check that the system is insulating at every point in $\mathbf{k}$-space, otherwise
the Berry phase is ill-defined. The relaxed ionic positions and self-consistent charge density are
then used as an input to the Berry phase calculation.

\begin{figure}[ht]
 \centering
 \includegraphics[width=0.4\columnwidth]{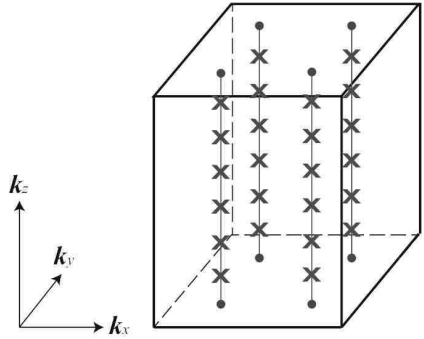}
 \caption{Choice of the $\mathbf{k}$-point grid for a Berry phase calculation of the polarization.
Here the polarization is to be calculated along the $z$-direction. The integration to obtain
the Berry phase is carried out along 4 strings of $\mathbf{k}$ points centered around $\Gamma$ 
in the $\mathbf{k}_x - \mathbf{k}_y$ plane, with 6 sampling points along each string in the
$\mathbf{k}_z$ direction. The final Berry phase is obtained by averaging the values obtained
from each of the four strings.
}
 \label{string_integration}
\end{figure}

\begin{figure}[ht]
 \centering
 \includegraphics[width=0.8\columnwidth]{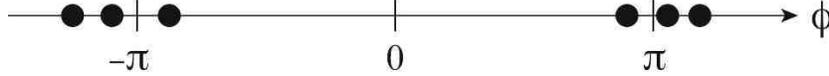}
 \caption{The black dots show the values of Berry phase obtained by integration along
six $\mathbf{k}$-point strings in the Brillouin zone. Clearly the average value is
close to $\pi$, $\pm$ the Berry phase quantum of $2\pi$. Mapping all of the values
into the lowest phase branch then taking the simple average would lead to an incorrect
result close to zero.
}
\label{kptaverage}
\end{figure}

One then proceeds to calculate one of the Berry phase values on the right-hand side
of Eqn.~\ref{P_BP}, that is
\begin{equation}
\sum_n^{occ} \int_{BZ} d^3 \mathbf{k} e^{-i\mathbf{k.R}} 
 \left< u_{n \mathbf{k}}| \frac{\partial u_{n \mathbf{k}}}{\partial \mathbf{k}} \right>
\quad ,
\end{equation}
where $u_{n \mathbf{k}}$ is the cell part of the Bloch function for the structure we are 
considering. First, the matrix elements are calculated by integrating along strings of $k$-points.
Since $ \frac{\partial}{\partial \mathbf{k}}$ is a vector derivative the matrix elements should be 
computed along any three non-collinear directions; usually the lattice vectors are chosen.
Then multiple strings in a particular direction are sampled so that an integration over the 
Brillouin zone can be performed (see Figure~\ref{string_integration}). 
It's important to check convergence both with respect to the number of
$k$-points along a string, and the number of strings used in the sampling, as the requirements
can be quite different in each case. Finally the values for all bands $n$ are summed. 
One subtletly, which is sometimes not well taken care of in codes, concerns the procedure for
averaging the Berry phase over the Brillouin zone. This is usually done by taking the sum of the
Berry phase values at each $k$-point, weighted by the fractional contribution of the $k$-point.
This procedure works well provided that the value from each $k$-point is on the same branch
of the polarization lattice. Figure~\ref{kptaverage} illustrates a not-uncommon problem that
can occur with some codes during the averaging procedure. Here by inspection the average Berry 
phase is clearly close to $\pi$, modulo the phase quantum of $2\pi$. Taking a simple average
of the values mapped into the range between $\pm \pi$, however, would result in an incorrect
value close to zero. We recommend checking the values of Berry phase obtained for the individual
strings if your code performs an automatic averaging procedure!

For a spin-polarized system, the 
Berry phase calculation is performed for both up- and down-spin electrons separately; the
phases are converted into polarization units by multiplying by $-\frac{ie}{2\pi^3}$ and then
added to the ionic contribution $\frac{1}{\Omega}\Sigma_i q_i \mathbf{r}_i$, where $q_i$ is the
charge of the pseudopotential or ion, to obtain the total
polarization of the system along the chosen lattice vector.

The above procedure is repeated for each lattice vector in turn. Be careful to check in the 
output of your code whether the results are reported with respect to the lattice vectors or 
in cartesian coordinates!

\begin{figure}[ht]
 \centering
 \includegraphics[width=0.9\columnwidth]{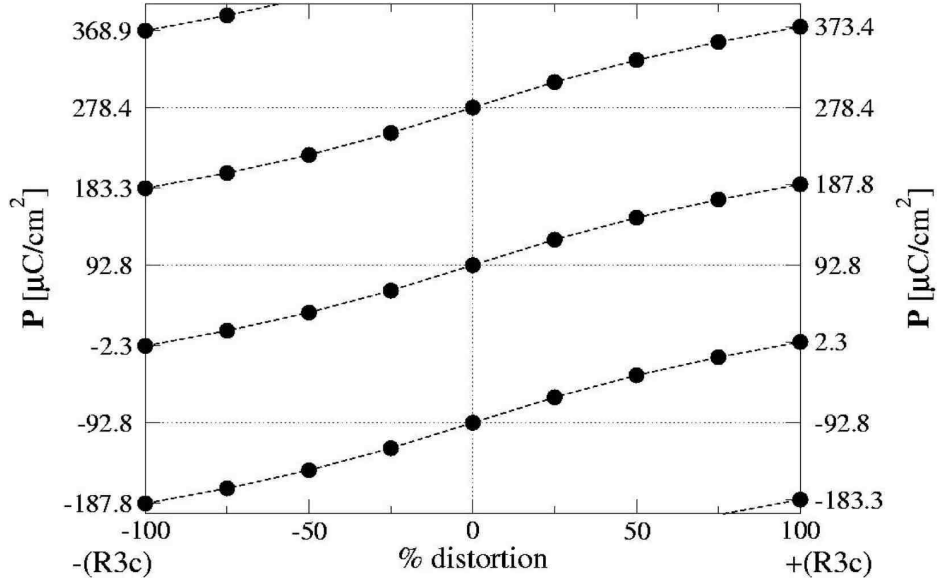}
 \caption{Calculated polarization as a function of percentage distortion from the
high symmetry non-polar structure (0\% distortion) to the ground state $R3c$ structure
for perovskite BiFeO$_3$. The black dots are calculated points and the dashed lines
are a guide to the eye illustrating the evolution along branches of the polarization
lattice. From Ref. \protect \cite{Neaton_et_al:2005}.
}
\label{BiFeO3}
\end{figure}

Remember that the number that you have now calculated is the absolute value of the polarization,
and is only defined modulo a polarization quantum. To calculate the spontaneous polarization
in a ferroelectric for example,
the procedure should be repeated also for a high symmetry, non-polar reference state. The 
difference between the two values, taken along the same branch of the polarization lattice, 
is then the spontaneous polarization. Sometimes it is necessary to re-calculate the polarization
for a number of structures along the deformation path between the high- and low-symmetry
structures in order to know unambiguously which difference to take. For example, Fig.~\ref{BiFeO3} 
shows the calculated polarization values for the case of perovskite structure BiFeO$_3$,
one of the most well-studied multiferroic materials\cite{Neaton_et_al:2005}. Notice first that 
the polarization lattice for the non-polar structure, labeled with 0\% distortion, does not 
contain zero, but is centered around 92.8 $\mu$C cm$^{-2}$, which is half a polarization quantum. 
It is clear from following the evolution of the polarization with distortion that the correct 
value for the spontaneous polarization is 187.8 - 92.8 = 95.0 $\mu$C cm$^{-2}$. From a calculation
of only the end-points at the $R3c$ and $-R3c$ structures the appropriate difference to take 
would be unclear, and one might incorrectly assume a value of $\frac{1}{2}(2.3 - (-2.3)) = 
2.3 \mu$C cm$^{-2}$. 
 
Finally a hint for calculating Born Effective Charges. Since these are defined as derivatives,
in principle the polarization should be calculated for the structure of interest, and then again
for an infinitesimally small displacement of each ion in turn. In practice, however, if the
diplacement is too small the result from this approach can be noisy. The best plan is to plot 
polarization as a
function of ionic displacement, starting with very small displacement values, and to take
the slope of the line in the region beyond the noise but before the non-linear regime.

\section{Last words}
I hope that this article has taken away some of the mystique associated with the modern
theory of polarization, and motivated you to start making your own calculations of
spontaneous polarization and related dielectric properties. For more practical introductory 
help, I recommend working through the tutorials that accompany many of the electronic structure
computational packages. For example the {\it Lesson on polarization and finite electric field}
provided by the ABINIT code, {\tt www.abinit.org}, is particularly helpful. Or even better,
attend a hands-on course hosted by one of the public codes where you will have direct access
to leading experts in the field. Good luck!
 
\section{Acknowledgements}
My thanks to the pioneers of the Modern Theory of Polarization -- Raffaele Resta and David
Vanderbilt -- who helped me to understand their elegant theory, as well as to the many students 
who have allowed me to impose my explanations upon them and in turn improve my description. 
The preparation of the manuscript was supported by ETH Z\"{u}rich.

\bibliographystyle{model1a-num-names}
\bibliography{/Users/nspaldin/papers/Nicola}

\end{document}